\documentclass{article}

\usepackage{arxiv}

\usepackage[utf8]{inputenc} 
\usepackage[T1]{fontenc}    
\usepackage{hyperref}       
\usepackage{url}            
\usepackage{booktabs}       
\usepackage{amsfonts}       
\usepackage{nicefrac}       
\usepackage{microtype}      
\usepackage{cleveref}       
\usepackage{subfig}         
\usepackage{lipsum}         
\usepackage{graphicx}
\usepackage{natbib}
\usepackage{doi}

\title{\emph{The World As Large Language Models See It}\thanks{The title refers to Albert Einstein's \textit{The World as I See It}.}

Exploring the reliability of LLMs in representing geographical features}

\newif\ifuniqueAffiliation
\uniqueAffiliationtrue

\ifuniqueAffiliation 
\author{{Omid Reza Abbasi} \\
	Department of Geoinformatics (Z\_GIS)\\
	Paris-Lodron University Salzburg\\
	Salzburg, Austria \\
	\texttt{omidreza.abbasi@plus.ac.at} \\
	\And
	{Franz~Welscher} \\
	Department of Geoinformatics (Z\_GIS)\\
	Paris-Lodron University Salzburg\\
	Salzburg, Austria \\
	\texttt{franz.welscher@plus.ac.at} \\
        \And
	{Georg Weinberger} \\
	Department of Geoinformatics (Z\_GIS)\\
	Paris-Lodron University Salzburg\\
	Salzburg, Austria \\
	\texttt{georg.weinberger@plus.ac.at} \\
        \And
	{Johannes Scholz} \\
	Department of Geoinformatics (Z\_GIS)\\
	Paris-Lodron University Salzburg\\
	Salzburg, Austria \\
	\texttt{johannes.scholz@plus.ac.at} \\
}
\else
\usepackage{authblk}

\newbox{\orcid}\sbox{\orcid}{\includegraphics[scale=0.06]{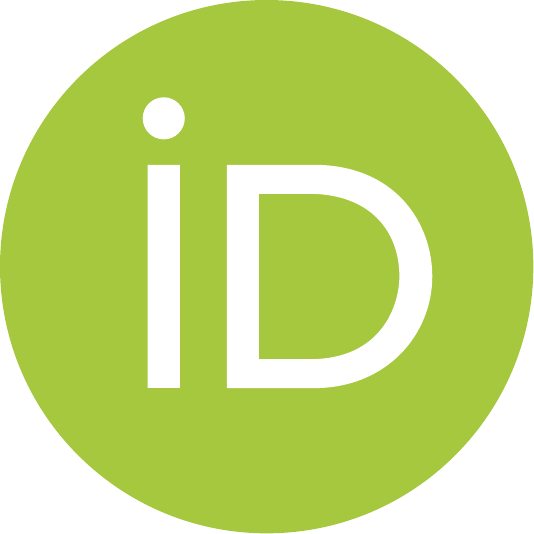}} 
\author[1]{%
	{Omid Reza Abbasi\thanks{\texttt{omidreza.abbasi@plus.ac.at}}}%
}
\author[1]{%
	{Franz Welscher\thanks{\texttt{franz.welscher@plus.ac.at}}}%
}
\author[1]{%
	{Georg Weinberger\thanks{\texttt{georg.weinberger@plus.ac.at}}}%
}
\author[1]{%
	{Johannes Scholz\thanks{\texttt{johannes.scholz@plus.ac.at}}}%
}
\affil[1]{Department of Geoinformatics (Z\_GIS)\\
	Paris-Lodron University Salzburg\\
	Salzburg, Austria}

\fi

\renewcommand{\headeright}{}
\renewcommand{\undertitle}{}
\renewcommand{\shorttitle}{\textit{The World As Large Language Models See It}}

\hypersetup{
pdftitle={A template for the arxiv style},
pdfsubject={q-bio.NC, q-bio.QM},
pdfauthor={Omid Reza Abbasi, Franz Welscher, Georg Weinberger, Johannes Scholz},
pdfkeywords={Large Language Models, Geographic Information Retrieval, AI Trustworthiness},
}

\begin{document}

\maketitle

\begin{abstract}
As large language models (LLMs) continue to evolve, questions about their trustworthiness in delivering factual information have become increasingly important. This concern also applies to their ability to accurately represent the geographic world. With recent advancements in this field, it is relevant to consider whether and to what extent LLMs' representations of the geographical world can be trusted. This study evaluates the performance of GPT-4o and Gemini 2.0 Flash in three key geospatial tasks: geocoding, elevation estimation, and reverse geocoding. In the geocoding task, both models exhibited systematic and random errors in estimating the coordinates of St. Anne's Column in Innsbruck, Austria, with GPT-4o showing greater deviations and Gemini 2.0 Flash demonstrating more precision but a significant systematic offset. For elevation estimation, both models tended to underestimate elevations across Austria, though they captured overall topographical trends, and Gemini 2.0 Flash performed better in eastern regions. The reverse geocoding task, which involved identifying Austrian federal states from coordinates, revealed that Gemini 2.0 Flash outperformed GPT-4o in overall accuracy and F1-scores, demonstrating better consistency across regions. Despite these findings, neither model achieved an accurate reconstruction of Austria's federal states, highlighting persistent misclassifications. The study concludes that while LLMs can approximate geographic information, their accuracy and reliability are inconsistent, underscoring the need for fine-tuning with geographical information to enhance their utility in GIScience and Geoinformatics.
\end{abstract}

\section{Introduction}
\label{sec:typesetting-summary}

Generative Artificial Intelligence (GenAI), particularly Large Language Models (LLMs), have transformed the way humans interact with AI. They have driven advancements in text generation \citep{li2024pre}, task automation \citep{wen2024autodroid,gebreab2024llm}, and decision making \citep{liu2024dellma,li2024stride} across various industries. As these models become more sophisticated, discussions about their applications and limitations have become increasingly important \citep{shi2025know,huang2024trustllm}. To evaluate their effectiveness, studies have tested these models in specialized fields such as healthcare \citep{bedi2024systematic}, finance \citep{zhao2024revolutionizing}, and law \citep{guha2024legalbench}, assessing their ability to produce accurate and contextually relevant information.

As LLMs continue to be integrated into specialized research domains, the field of GIScience has begun exploring their potential to enhance spatial analysis and geographic data processing \citep{wang2024gpt}. A growing body of research has examined their applicability in areas such as geographic information retrieval \citep{haris2024exploring, manvi2023geollm}, spatial reasoning \citep{majic2024spatial}, and automated cartography \citep{holm2024geogpt, zhang2024mapgpt}. These studies aim to evaluate the effectiveness and reliability of LLMs in addressing complex geospatial challenges and their ability to complement traditional GIS methodologies. However, some fundamental questions remain:

\begin{itemize}
    \item How accurate, reliable, and comprehensive is an LLM's internal perception of the real geographical world?
    \item In other words, can LLMs be regarded as independent geospatial data generators?
\end{itemize}

To investigate this, we design a series of geospatial tasks to understand the representation of real world in LLMs and to examine their geographical knowledge. Specifically, we evaluate GPT-4o and 2.0 Flash in geocoding, elevation estimation, and reverse geocoding. We then assess its output based on criteria such as accuracy and precision. Finally, we discuss the broader implications of our findings, highlighting key challenges related to the trustworthiness and reproducibility of LLM-generated spatial data.

\section{Related works}

Recent advances in LLMs have spurred interest in their ability to serve as repositories of factual knowledge. \citet{brown2020language} demonstrated that LLMs can perform a wide range of tasks in a few-shot setting, hinting at their broad potential. \citet{petroni2019language} further investigated the notion of “language models as knowledge bases,” showing that these models store a surprising amount of factual information. However, a growing body of work has raised concerns regarding the reliability of LLM outputs. For instance, \citet{bender2021dangers} argued that LLMs, due to their reliance on statistical patterns, are prone to generating “hallucinated” or biased content. More recently, \citet{wei2022emergent} documented emergent abilities of LLMs while also noting systematic limitations, especially when models are queried for detailed or domain-specific facts.

Recent studies have evaluated LLMs' proficiency in spatial reasoning. For instance, \citet{yamada2023evaluating} study assessed GPT-3.5 and GPT-4 on tasks involving navigation through various spatial structures, including grids and trees. The findings revealed variability in performance across different spatial configurations, indicating that while LLMs can grasp certain spatial concepts, their understanding is not yet comprehensive. Generative AI models have been increasingly utilized to automate geospatial data handling \citep{li2023autonomous, ning2024autonomous, mansourian2024chatgeoai}. For instance, a tool called MapGPT \citep{zhang2024mapgpt} has been developed to simplify mapping and geospatial analysis, making the processes more accessible to both novices and experts. Although these studies have advanced automated analysis of geospatial data, they rely on code generation mechanisms in LLMs. Our study differs from these by leveraging the internal knowledge of LLMs to generate geospatial data.

The aim of our study is more closely aligned with works focused specifically on LLMs' understanding of the geographical world. For instance, \citet{o2024metric} investigated the spatial reasoning capabilities of LLMs, revealing both inherent knowledge and notable limitations. Their findings suggest that while LLMs demonstrate an understanding of geospatial concepts such as geo-coordinates and the terms \textit{near} and \textit{far}, they struggle to adjust the scale of these terms appropriately within different contexts. \citet{bhandari2023large} investigated the geospatial knowledge, awareness, and reasoning capabilities of LLMs through a series of experiments. The study evaluated LLMs' ability to represent geographic coordinates, interpret geospatial prepositions, and infer city locations using multidimensional scaling. The findings highlight that while larger and more advanced models demonstrate improved geospatial understanding, significant limitations remain in their ability to synthesize and apply geographic knowledge effectively. \citet{liu2024measuring} examined the geographic diversity of GPT-4 as a knowledge base by assessing its representation of geographic features through a geo-guessing experiment. Using DBpedia as ground truth, the study reveals that GPT-4 exhibits insufficient knowledge of certain geographic feature types and inter-regional disparities in performance, particularly on UNESCO World Heritage Sites. The findings highlight variations between GPT-4's unimodal and multimodal versions and call for further discussion on geographic diversity as an ethical principle in GIScience. \citet{roberts2023gpt4geo} explored GPT-4’s ability to acquire and apply factual geographic knowledge through a series of experiments ranging from basic location and distance estimation to complex tasks like route finding and supply chain analysis. The study assesses the model’s strengths and limitations in handling geographic data without external tools or internet access. The findings provide insights into GPT-4’s potential for geospatial applications while highlighting areas where its reasoning and accuracy remain constrained.

\section{Evaluating LLMs' geospatial knowledge}

In this section, we define three geospatial tasks to evaluate the performance of GPT-4o and the Gemini 2.0 Flash model in responding to queries. For each task, we provide a brief description and the prompt used to generate the response. The results are then analyzed, and their implications are discussed. To interact with the models, we used the OpenAI API and the Gemini API. All queries were performed using zero-shot prompts.

\subsection{Geocoding}

Converting place names into geographical coordinates, known as geocoding, involves matching location names to latitude and longitude pairs by querying large-scale geographic databases such as GeoNames\footnote{https://www.geonames.org/} or using geocoding services such as the Google Maps Geocoding API\footnote{https://developers.google.com/maps/documentation/geocoding} or Nominatim API\footnote{https://nominatim.org/}. In this task, we asked the model to estimate the location of St. Anne's Column (Annasäule) in Innsbruck, Austria (47.265556 N, 11.394167 E). We purposely selected a point feature to avoid concerns related to ambiguous boundaries in areal features. To measure the accuracy and precision of the estimations, we repeated the task 100 times.

\begin{description}
\item[Prompt 1] What are the geographical coordinates of St. Anne's Column in Innsbruck, Austria? Provide the coordinates with 6 digits after decimal.
\end{description}

Figure 1 shows the estimations of GPT-4o and 2.0 Flash, along with the true location of St. Anne's Column. A statistical analysis of the estimated coordinates by GPT-4o reveals both systematic and random errors in the measurements. The average of the estimates was (47.26749 N, 11.39469 E), indicating a systematic northward bias of approximately 214 m and an eastward bias of approximately 39 m. After removing this bias, the precision of the measurements was quantified by standard deviations of roughly 153 m in latitude and 32 m in longitude, yielding an overall RMSE of 268 m. Furthermore, the error covariance matrix was computed. An eigenanalysis of this matrix indicated that the 1–\( \sigma \) error ellipse has semi–axes of about 154 m and 29 m, with its major axis oriented approximately 5° east of north. When scaled to the 95\% confidence level, these dimensions correspond to semi-axes of ellipse of roughly 376 m and 72 m, respectively.

For 2.0 Flash, our analysis again indicates a systematic error in the north–south direction. In this case the mean estimated position is approximately (47.26840, 11.39417), which implies a mean latitude bias of about 316 m northward, while the mean longitude is virtually unchanged (about 2–3 m eastward). In contrast, the precision  component of the error is much smaller. After removing the bias, the standard deviations are on the order of 18.5 m in latitude and 15 m in longitude. Consequently, the RMSE of an individual measurement relative to the true point is approximately 317 m, dominated by the systematic offset in latitude. The corresponding 1–\( \sigma \) error ellipse has semi–axes of roughly 18.5 m in the north–south direction and 15 m in the east–west direction. At the 95\% confidence level, the error ellipse expands to semi–axes of about 45 m and 37 m, respectively. The results shows that 2.0 Flash's estimations are more precise but significantly offset from the true location, with the systematic error in latitude being the dominant contributor to the overall RMSE.

A visual analysis of the study area around the estimated locations suggests that the systematic bias is likely due to the distribution of other landmarks in the vicinity. Specifically, the Goldenes Dachl (Golden Roof), Hofkirche (Court Church), Stadtturm (City Tower), and several other historical landmarks are situated to the north of St. Anne's Column. It appears that these prominent locations are pulling the estimations toward them.

\begin{figure}[!h]
    \centering
    \includegraphics[width=1\linewidth]{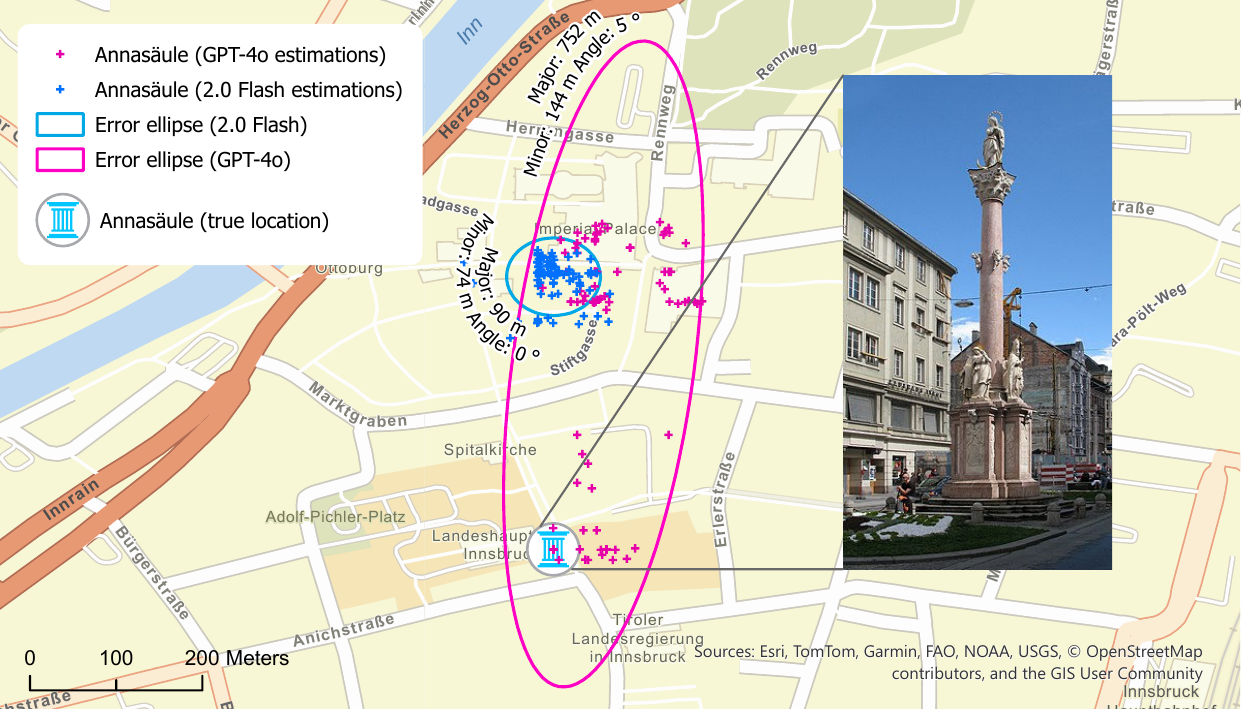}
    \caption{The true location of St. Anne’s Column in Innsbruck, Austria, along with estimated locations and the corresponding error ellipse. (Image credit: Andrew Bossi, CC BY-SA 2.5, via Wikimedia Commons.)}
    \label{fig:enter-label}
\end{figure}

\subsection{Elevation estimation}

Digital Elevation Models (DEMs) are generated using remote sensing techniques such as LiDAR, photogrammetry, and radar to measure the Earth's surface elevations with high precision. Although LLMs can analyze textual descriptions and metadata related to DEMs, they do not inherently understand or process elevation data in the same way as specialized GIS software. In this task, we examined the ability of models to estimate the elevations of a grid of coordinates across Austria, spaced 10 km apart (840 points in total). The following prompt was used to extract elevation information from the models.

\begin{description}
\item[Prompt 2] Tell me the elevation of this coordinate ({lat} N, {lon} E) in meters above sea level?
\label{item:p1}
\end{description}

We repeated the task five times to reduce noise in the responses. For each cell, the median of the estimates was then used to compute the DEM, a choice made because the median is less affected by extreme values, especially those observed in the highland regions of the Alpine area. Figure 2(a) shows the DEM of Austria used as ground truth dataset\footnote{Data provided by data.gv.at, downloaded from \href{https://www.data.gv.at/katalog/en/dataset/d88a1246-9684-480b-a480-ff63286b35b7}{https://www.data.gv.at/katalog/en/dataset/d88a1246-9684-480b-a480-ff63286b35b7} and is licensed under \href{https://creativecommons.org/licenses/by/4.0/deed.de}{Creative Commons Namensnennung 4.0 International (CC BY 4.0)}.}.The official DEM was originally provided at a 10-meter resolution. To align with our grid, we aggregated the values using the median to achieve a 10 km-resolution version. Figure 2(b) and 2(d) present the DEM generated from GPT-4o and 2.0 Flash's responses. Figure 2(c) and Figure 2(e) depict the relative errors (residuals) in the estimated elevations. A one-sample \textit{t}-test was conducted to assess whether the mean difference between the official and estimated elevation data deviated significantly from zero. Table 1 summarizes the results of the \textit{t}-test statistic.

\begin{figure}[!h]
    \centering
    \includegraphics[width=1\linewidth]{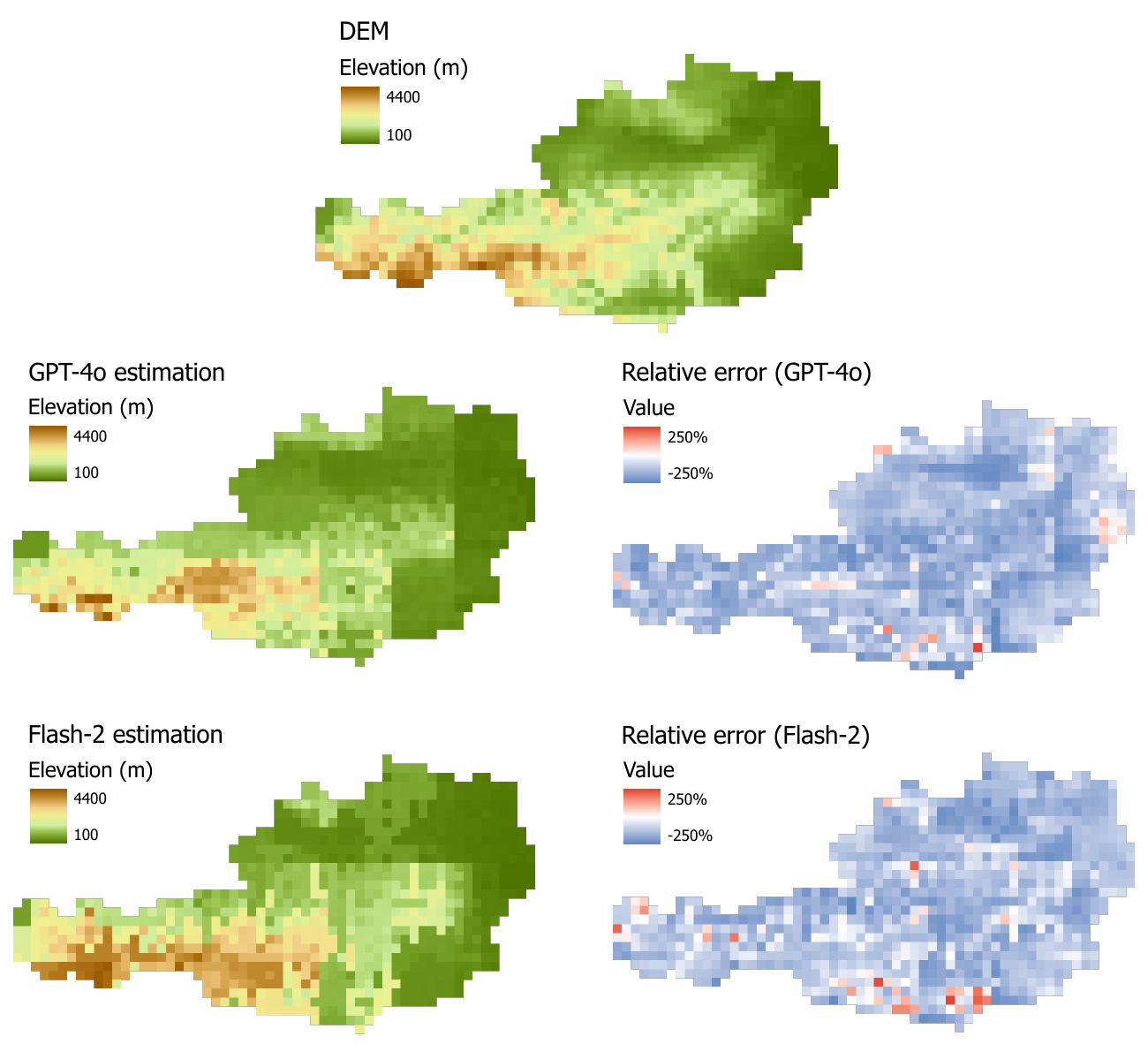}
    \caption{(a) DEM of Austria; (b) Estimated DEM generated from GPT-4o responses; (c) Relative errors of the estimations by GPT-4o; (d) Estimated DEM generated from 2.0 Flash responses; (e) Relative errors of the estimations by 2.0 Flash.}
    \label{fig:dem}
\end{figure}

\begin{table}[!h]
\centering
\caption{Summary of statistical test results for GPT-4o and 2.0 Flash models}
\label{tab:stats}
\begin{tabular}{lcccccc}
\hline
\textbf{Model} & \textbf{Mean} & \textbf{Std. Dev.} & \textbf{Min} & \textbf{Max} & \textbf{\textit{t}-statistic} & \textbf{\textit{p}-value} \\
\hline
\textbf{GPT-4o} & 114.26 & 343.77 & -1239.93 & 1314.44 & 10.32 & $9.38\times10^{-24}$ \\
\textbf{2.0 Flash} & 43.51 & 393.82 & -1780.84 & 1764.85 & 3.43 & 0.00063 \\
\hline
\end{tabular}
\end{table}
 
In the case of GPT-4o, the average difference between the official and estimated elevations is approximately 114.26 meters. Since the mean is significantly different from zero, it suggests that, on average, the model is underestimating the official elevation by this amount. The \textit{t}-statistic indicates that the observed mean difference is more than 10 standard errors away from zero, and the extremely low \textit{p}-value (far below the common significance threshold of 0.05) confirms that this result is highly unlikely to be due to random chance. The standard deviation indicates that the individual errors vary quite a bit around the mean. Although there is a considerable spread in the residuals, the systematic bias remains evident when considering the average. In addition, some errors are much greater than the average error in magnitude (both negative and positive). This suggests that while a systematic bias is present, local variability or outliers also contribute to the overall error distribution.

In the case of 2.0 Flash model, the average difference between the official and the estimated elevations is about 43.51 meters. The \textit{t}-statistic indicates that the mean difference is approximately 3.43 standard errors away from zero. The \textit{p}-value is well below the typical significance threshold (0.05), providing strong evidence against the null hypothesis. In other words, the observed mean difference is statistically significant, indicating that the systematic bias in the 2.0 Flash model’s estimates is unlikely to be due to random chance. Although a bias is present in the estimations, its magnitude is notably lower than what was observed with the GPT-4o model. The wide range of residuals suggests that although the average bias is moderate, some individual estimates deviate significantly from the official values. This observation further underscores the variability in the model’s performance across the study area. Figure 2 shows that both models tend to underestimate elevations across most of the study area, yet they capture the overall topographical trends. The 2.0 Flash model performs better in the eastern regions compared to GPT-4o; however, it occasionally overestimates elevations at specific points.

\subsection{Reverse geocoding}

Reverse geocoding is the process of converting geographic coordinates (such as latitude and longitude) into a human‐readable address or place name by querying spatial databases and applying geospatial algorithms \citep{longley2015geographic}. It plays a critical role in applications such as location-based services, navigation systems, and emergency response, where its precision is highly dependent on the quality and granularity of the underlying geospatial data \citep{haklay2008openstreetmap}. In our study, we explored how well LLMs can identify the Austrian federal states. The same grid of 840 points used in the first task was also employed here. First, we asked the models to assign a federal state to each point based solely on its coordinates. We repeated the task five times and picked the most frequent value. The following prompt was used to extract state information from the models.

\begin{description}
\item[Prompt 3] Tell me in which state of Austria this coordinate ({lat} N, {lon} E) is located?
\end{description}

\begin{figure}[!h] 
    \centering
    \subfloat[The confusion matrix resulting from the reverse geocoding task conducted by GPT-4o.]{%
        \includegraphics[width=0.48\textwidth]{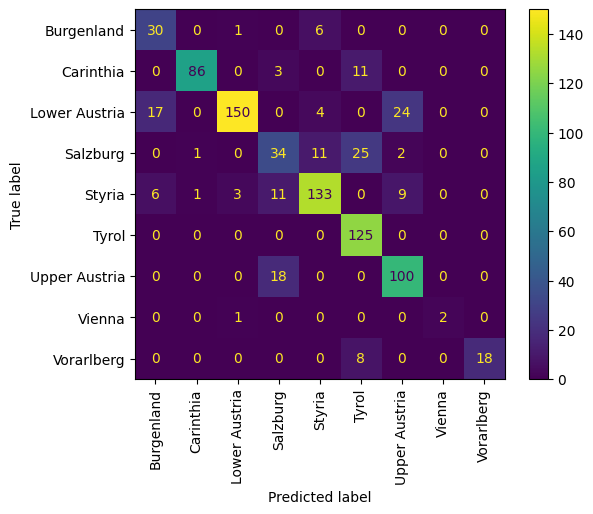}%
        \label{fig:confusion_gpt}%
        }%
    \hfill%
    \subfloat[The confusion matrix resulting from the reverse geocoding task conducted by 2.0 Flash.]{%
        \includegraphics[width=0.48\textwidth]{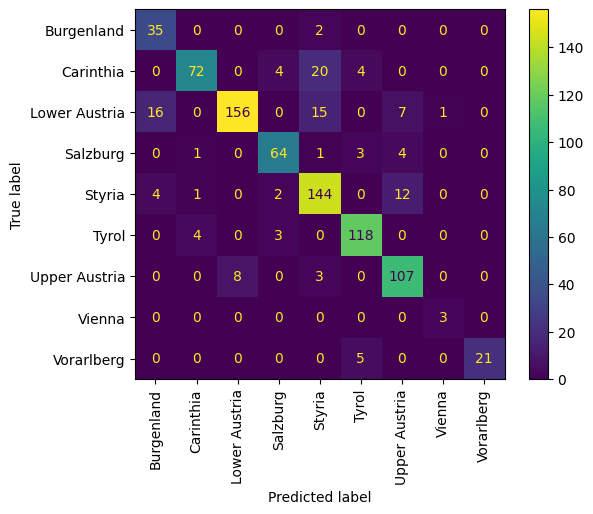}%
        \label{fig:confusion_flash}%
        }%
    \caption{The confusion matrices of the reverse geocoding task}
    \label{fig:confusion}
\end{figure}

To evaluate the performance, we compared the estimations with the results of the Nominatim reverse geocoding service as a benchmark. We visualized the agreement and discrepancies between the two sets of results using a confusion matrix. To gain a more detailed understanding, we also calculated key classification metrics such as precision, recall, and F1-score. Figure \ref{fig:confusion} shows the confusion matrix resulting, in which the labels are the nine federal states of Austria. In addition, Table 2 summarizes the metrics for the accuracy and precision of the predictions.

\begin{table}[!h]
    \centering
    \small
    \setlength{\tabcolsep}{5pt}
    \renewcommand{\arraystretch}{1.2}
    \caption{Comparison of GPT-4o and Gemini 2.0 Flash for reverse geocoding in Austria}
    \begin{tabular}{lcccccccc}
        \hline
        \textbf{Region} & \multicolumn{3}{c}{\textbf{GPT-4o}} & \multicolumn{3}{c}{\textbf{Gemini 2.0 Flash}} & \textbf{Support} \\
        \hline
        & Precision & Recall & F1-score & Precision & Recall & F1-score & \\
        \hline
        Burgenland      & 0.57 & 0.81 & 0.67 & 0.64 & 0.95 & 0.76 & 37  \\
        Carinthia       & 0.98 & 0.86 & 0.91 & 0.92 & 0.72 & 0.81 & 100 \\
        Lower Austria   & 0.97 & 0.77 & 0.86 & 0.95 & 0.80 & 0.87 & 195 \\
        Salzburg        & 0.52 & 0.47 & 0.49 & 0.88 & 0.88 & 0.88 & 73  \\
        Styria         & 0.86 & 0.82 & 0.84 & 0.78 & 0.88 & 0.83 & 163 \\
        Tyrol           & 0.74 & 1.00 & 0.85 & 0.91 & 0.94 & 0.93 & 125 \\
        Upper Austria   & 0.74 & 0.85 & 0.79 & 0.82 & 0.91 & 0.86 & 118 \\
        Vienna          & 1.00 & 0.67 & 0.80 & 0.75 & 1.00 & 0.86 & 3 \\
        Vorarlberg      & 1.00 & 0.69 & 0.82 & 1.00 & 0.81 & 0.89 & 26  \\
        \hline
        \textbf{Accuracy} & \multicolumn{3}{c}{0.81} & \multicolumn{3}{c}{0.86} & 840 \\
        \textbf{Macro avg} & 0.82 & 0.77 & 0.78 & 0.85 & 0.88 & 0.85 & 840 \\
        \textbf{Weighted avg} & 0.83 & 0.81 & 0.81 & 0.87 & 0.86 & 0.86 & 840 \\
        \hline
    \end{tabular}
\end{table}

In the case of the GPT-4o model, the overall accuracy of 81\% suggests that the model is fairly reliable in determining the correct region, though the performance varies across different Austrian states. Some regions, such as Vorarlberg, show perfect precision but relatively low recall. This suggests that while the model rarely misclassifies other regions as Vorarlberg, it often fails to correctly assign points that actually belong to this region. Carinthia and Styria exhibit strong balance between precision and recall, indicating consistent reliability in identifying these regions. Salzburg shows the lowest scores, suggesting significant difficulty in distinguishing Salzburg from other regions. Burgenland also has relatively low precision, meaning many points classified as Burgenland actually belong to other states. The macro-averaged F1-score (0.78), which treats all regions equally, is slightly lower than the weighted average F1-score (0.81), which accounts for class imbalance. This suggests that GPT-4o performs better on frequently occurring regions, but struggles more with less common ones.

In the case of 2.0 Flash, the model achieves an overall accuracy of 86\%, indicating better performance in identifying the correct regions. Several regions exhibit both high precision and high recall, indicating that the model consistently assigns the correct labels. Notable examples include Salzburg and Tyrol. Compared to GPT-4o, Burgenland and Upper Austria show a significant increase in recall, meaning fewer instances were incorrectly classified as other states. Carinthia and Vorarlberg have relatively lower recall scores, suggesting that some points in these regions were misclassified into neighboring states. While Vienna has a high precision score, its recall has improved to 1.00 compared to GPT-4o’s 0.67 recall, meaning all Vienna points were correctly identified, though some other locations have been misclassified as Vienna. Gemini 2.0 Flash outperforms GPT-4o across all major metrics, with a 5\% higher accuracy and improved recall in most regions. The macro-averaged F1-score (0.85) and weighted F1-score (0.86) show a notable improvement over GPT-4o, suggesting that Gemini 2.0 Flash performs more consistently across all classes and handles class imbalances better. GPT-4o struggled with Salzburg (0.49 F1-score) and Burgenland (0.67 F1-score), while Gemini 2.0 Flash improved to 0.88 and 0.76, respectively.

\begin{figure}[h!] 
    \centering
    \subfloat[The map of federal states of Austria]{%
        \includegraphics[width=0.90\textwidth]{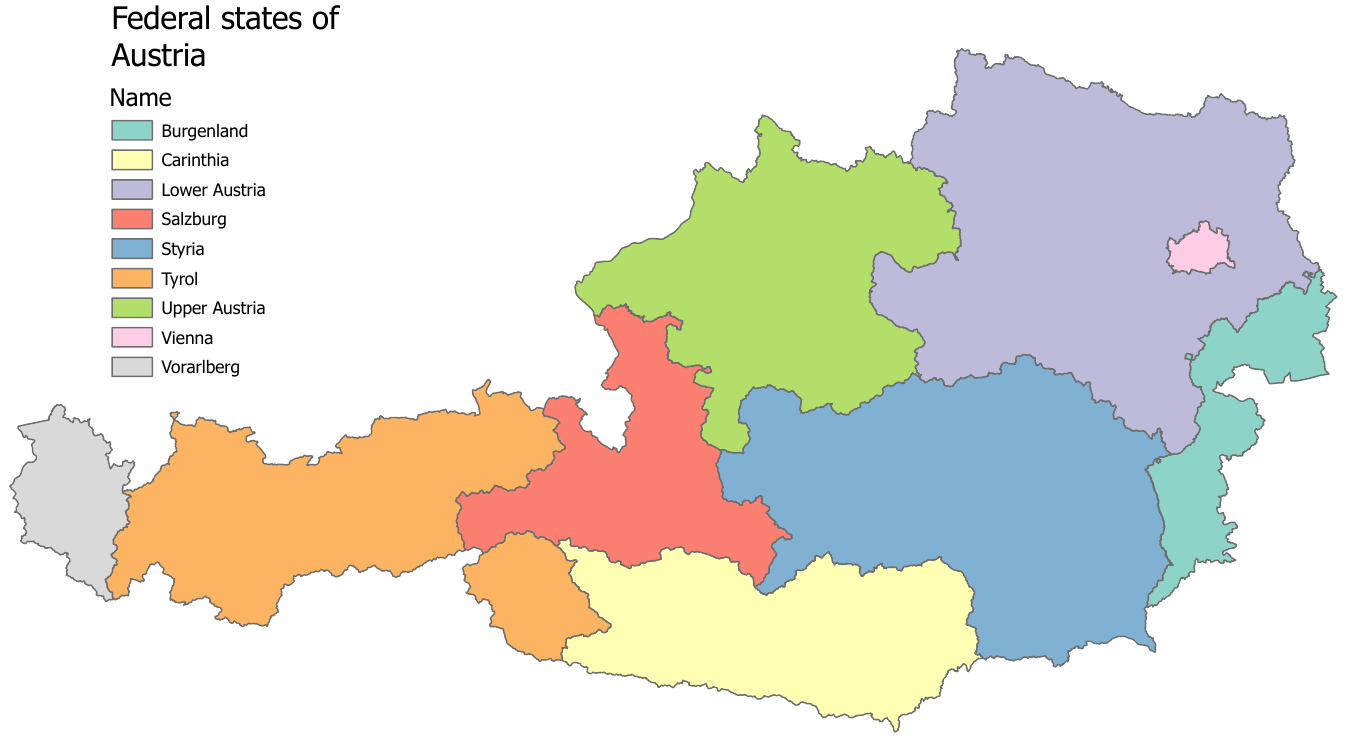}%
        \label{fig:states_official}%
        }%
    \hfill%
    \subfloat[The map of federal states of Austria as perceived by the GPT-4o model]{%
        \includegraphics[width=0.45\textwidth]{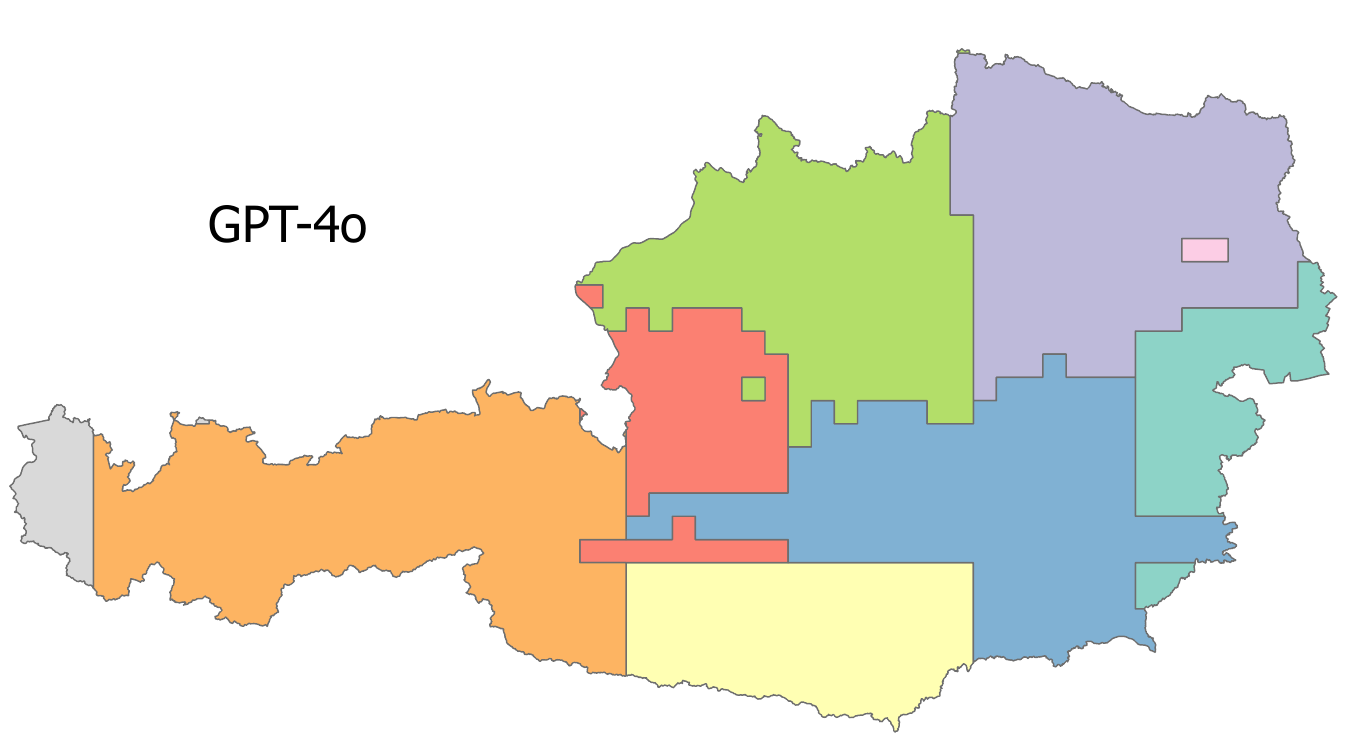}%
        \label{fig:states_gpt}%
        }%
    \hfill%
    \subfloat[The map of federal states of Austria as perceived by the 2.0 Flash model]{%
        \includegraphics[width=0.45\textwidth]{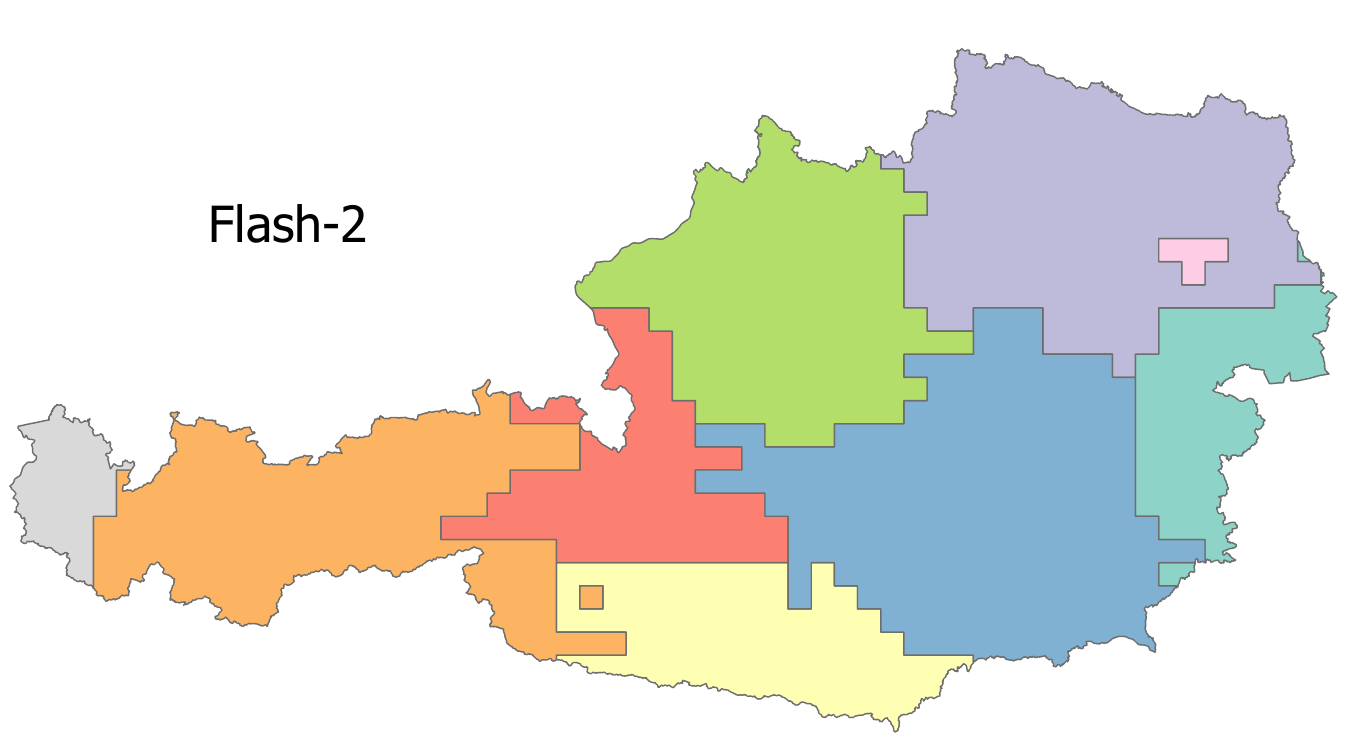}%
        \label{fig:states_flash}%
        }%
    \caption{The layout of Austria's states as perceived by the models in the task of reverse geocoding}
    \label{fig:states}
\end{figure}

Figure 4 sketches a layout of Austria's states as perceived by the models in the task of reverse geocoding. While both models attempt to reconstruct Austria’s federal states, neither achieves an accurate representation. 2.0 Flash’s output retains a better overall shape and organization of Austria’s states compared to GPT-4o. This is more apparent in Salzburg where GPT-4o fails to identify substantial parts of the state.

\section{Conclusions}

Our evaluation of GPT-4o and Gemini 2.0 Flash in geospatial tasks reveals that while LLMs can approximate geographic information, their accuracy and reliability remain inconsistent. Both models exhibit systematic biases, with GPT-4o showing greater deviations in geocoding and elevation estimations, while Gemini 2.0 Flash provides more precise but still imperfect results. In reverse geocoding, Gemini 2.0 Flash outperforms GPT-4o, yet misclassifications persist. These findings underscore the limitations of LLMs in answering geographical questions and highlight the importance of fine-tuning LLMs with geographical information. Future work should focus on improving LLMs’ geospatial reasoning and mitigating systematic biases to enhance their usability in GIScience and Geoinformatics applications. 


\bibliographystyle{apalike}
\bibliography{article}

\end{document}